**Non-randomness of Japan megaquakes implied by stress recovery and accumulation**

Kazuyoshi Z. Nanjo,[1,2,3,4,5]*, Takane Hori[5,1], Daichi Iwata[6]

[1]Global Center for Asian and Regional Research, University of Shizuoka, 3-6-1, Takajo, Aoi-ku, Shizuoka 420-0839, Japan.
[2]Department of Physics and Astronomy, University of California, Davis, One Shields Avenue, Davis, CA 95616, USA.
[3]Center for Integrated Research and Education of Natural Hazards, Shizuoka University, Shizuoka, 836, Ohya, Suruga-ku, Shizuoka 422-8529, Japan.
[4]Institute of Statistical Mathematics, 10-3 Midori-cho, Tachikawa, Tokyo 190-8562, Japan.
[5]Yokohama Institute for Earth Sciences, Japan Agency for Marine-Earth Science and Technology, 3173-25, Showa-machi, Kanazawa-ku, Yokohama, Kanagawa 236-0001, Japan.
[6]Independent researcher, 3 Shirako, Wako, Saitama 351-0101, Japan.

*Correspondence and requests for materials should be addressed to K.Z.N. (email: nanjo@u-shizuoka-ken.ac.jp); Tel.: +81-54-245-5600

ORCIDs
https://orcid.org/0000-0003-2867-9185 (K. Z. Nanjo)
https://orcid.org/0000-0003-3769-3894 (T. Hori)

Emails
nanjo@u-shizuoka-ken.ac.jp (K. Z. Nanjo)
horit@jamstec.go.jp (T. Hori)
dai.iwata.r@gmail.com (D. Iwata)

**Abstract**

Monitoring stress recovery and accumulation associated with megaquakes helps to assess their recurrence. Previous studies proposed a high likelihood of imminent recurrence for the 2011 Tohoku and the 17th-century Hokkaido megaquakes belonging to the magnitude-9 class, although their current stress state remains uncertain. Here we compare the occurrence of small earthquakes relative to larger ones, using *b*-values, showing high *b*-values in the source area of the Tohoku earthquake, indicating low stresses. In contrast, low *b*-values occurred in the source area of the 17th-century earthquake, indicating high stresses as seen before the Tohoku event. Around the low-*b*-value zone, phenomena that provide insight into subsequent large earthquakes are observed, and which were reported for worldwide earthquakes: seismic quiescence, a seismic gap, strong plate coupling, and slow earthquake activity avoiding megaquakes' rupture zones. Results imply that the Hokkaido and Tohoku megaquakes occur nonrandomly in time, rather their recurrence intervals are more characteristic.

## Introduction

Many large underthrust earthquakes have occurred off the Pacific coast of Japan. Examples include the 2011 Tohoku earthquake of magnitude (*M*) 9 off the Tohoku coast[1-4] and the 2003 *M*8 Tokachi earthquake off the Hokkaido coast[5]. Such earthquakes are thought to rupture plate interface faults to release the strain energy and relax stress that accumulates due to plate coupling. However, from recent observations in a section of the subducting Pacific plate beneath Tohoku and Hokkaido districts, it is still uncertain whether sufficient energy remains in the system to allow similar size events more or less immediately. Those observations showed that a drop in stress during the Tohoku earthquake was large enough, relative to the background stress, near the high-slip zone of that earthquake[6-8]. In contrast, another study showed that the stress relieved by the earthquake recovered very rapidly to levels before the earthquake[9]. For the 2003 *M*8 Tokachi earthquake, sufficient energy remained near the high-slip zone of the earthquake[7,10,11]. Moreover, previous studies involving paleotsunami research revealed that a great earthquake with a comparative magnitude to the 2011 Tohoku event occurred off Hokkaido in the 17th century[12-14], causing widespread tsunamis, although the Tokachi earthquake did not generate such widespread tsunamis. The Headquarters for Earthquake Research Promotion stated a high likelihood of its imminent recurrence, given the recurrence interval of 17th-century-type earthquakes and the occurrence time of the latest event[15], where the recurrence interval and latest occurrence time were estimated using tsunami deposits. Geophysical monitoring along the plate interface off the Japan Pacific coast, even after the occurrence of large earthquakes, is important for evaluating whether large earthquakes may have a characteristic recurrence interval or occur more randomly over time.

An approach for this type of monitoring is to use a fundamental observation in seismology: the *b*-value of the Gutenberg-Richter (GR) law[16], given as $\log_{10}N=a-bM$, where $N$ is the cumulative number of earthquakes with a magnitude larger than or equal to $M$, $a$ characterizes seismic activity or earthquake productivity of a region, and $b$ is used to describe the relative occurrence of large and small events (i.e., a high *b*-value indicates a larger proportion of small earthquakes, and vice versa). The insets in Fig. 1 show a fit of the GR law to observations in the present cases.

Laboratory experiments have shown that the *b*-value decreases with stress[17-19]. Moreover, there have been a number of observations that indicate that this relation also holds for earthquakes. For example, events on normal faults that tend to be under lower stress have the highest *b*-values, while events on thrust faults that tend to be under higher stress have the lowest *b*-values, where strike-slip events have intermediate *b*-values[20]. Using a frictional model for stresses in the continental lithosphere, it was found that *b* was inversely proportional to stress in a wide variety of tectonic regions[21]. There are indications of the role of *b* as a stressmeter from laboratory experiments[17] and numerical simulations[22]. One possible model showed that the decrease in *b* was caused by a rapid increase in shear stress, promoting micro-crack growth[17]. Another possible model showed that an increased stress level in slowly slipping parts of a fault facilitates earthquake rupture growth, resulting in the temporal decrease in *b* prior to a large earthquake (mainshock)[22]. As noted above, the role of *b* as a stressmeter is justified not by a single study, but by multiple studies. Moreover, decreases in *b* and/or low *b*-values (indicative of stress increase and/or high stress) have been reported for the period leading up to large earthquakes in Sumatra (Indonesia), Japan, Chile, California (USA), and Italy[9,23-27]. Within the framework of the earthquake-cycle theory that combines the stress-accumulation hypothesis and the elastic-rebound theory[28,29], *b*-values may help to assess the stress state of a fault (i.e., interseismic, preseismic, or postseismic) on which earthquakes repeatedly occur.

Here, the characteristics of seismicity off the Japanese Pacific coast are investigated. This paper focuses on determining maps of *b*-values, with data of good enough quality and sufficient abundance, collected by the JMA (Japan Meteorological Agency) (for details on seismicity dataset, see Methods). Since the JMA earthquake catalog is used, *M* indicates the JMA magnitude. The local variation in *b*-values shows that an area of low-and-deceasing *b*-values off Hokkaido is linked with phenomena that provide insight into future large earthquakes. Namely, they are strong coupling of the plate interface, estimated from crustal deformation[30], seismic quiescence based on seismicity observed for about 60 years[31], slow earthquake activity that did not overlap with the megaquake's rupture zone, compiled from a previous study[32], and a seismic gap between magnitude-8-class megaquakes, estimated from historical records[5,33,34]. Moreover, from a comparison with a 17th-century earthquake[14], the low-and-decreasing *b*-value zone is thought to be in or near the margin of the high-slip area of this 17th-century earthquake. Based on our result, we conclude that megaquakes off the Japan Pacific coast may have a characteristic interval, although this does not support a recent hypothesis[9] that claims that megaquakes is randomly distributed over time.

## Results

**Spatiotemporal changes of *b*-values.** Our approach considered only seismicity along plate interface between the subducting Pacific plate and the overriding continental plate (see Methods and Supplementary Figs. 1-3). The outer rise, east of the trench axis, which exhibits lower differential stress typical of a shallow normal-type faulting regime, has no continental plate sitting atop it. We truncated the seismicity catalog to discard in advance all data east of the trench axis. Because coseismic-slip areas of large earthquakes on the plate boundary mostly were shallower than ~60 km (Fig. 1 and Supplementary Fig. 4), we focused on *b*-values shallower than ~60 km.

A map view based on seismicity along the plate interface shows that the *b*-values in the large co-seismic slip zone of the 2011 *M*9 Tohoku earthquake[35] were distinctly low before the earthquake (Fig. 1a)[9,23]. For details on *b*-value analysis, see Methods. After the Tohoku earthquake (Fig. 1b), the *b*-values were high in the co- and post-seismic slip zones. North and south from these zones, the *b*-values were still low (Fig. 1b). We considered three post-Tohoku periods to show that the pattern remained stable (Supplementary Fig. 5).

For earthquakes in the circle indicated by 1 in Fig. 1, the *b*-values show a gradual decrease over time, to values near 0.6 before the 2011 Tohoku earthquake (Fig. 2a). This feature was observed in previous studies[9,23]. After the major earthquake, they remained at values near 1[36] (Fig. 2a). For details on *b*-value analysis, see Methods and Supplementary Fig. 6.

A previous study[9], which used data up to 2014, documented a sudden increase in the *b*-value at the time of the earthquake and its return to a decreasing trend in *b*-values.

Assuming that the *b*-value is negatively correlated with differential stress[17], this infers that the stress relieved by the earthquake recovered very rapidly to levels before the earthquake. However, our current study, which used data until Jun. 2025, showed no such indication of a decreasing trend in *b*. Moreover, a previous study[9] misinterpreted strong temporal heterogeneity in *b* for the early period from immediately after the Tohoku earthquake to 2014. Together with previous observations[6,7] showing that the drop in stress during the Tohoku earthquake was large enough, relative to the background stress near the high-slip zone, we agreed that the hazard of a great tsunami-producing earthquake off the coast of the Tohoku district is unlikely to be already almost as high as it was just before the 2011 event[4,8]. Many centuries of stress accumulation are required to rebuild stress and produce another great event[4,8].

As was previously suggested[9,23], we found persistently low *b*-values after the 2003 *M*8 Tokachi earthquake in and around its high-slip zone[5] (Fig. 1). Assuming a stress-and-*b* correlation[17], a previous study[9] suggested that although an *M*8 event occurred, the earthquake did not release a significant amount of overall stress that is continuously accumulating along that part of the Pacific plate on a large enough area to host megathrust earthquakes[10]. This is consistent with geodetic observations that reflect only minor fluctuations in the subsidence rate related to *M*7~8 earthquakes over the past 120 years[11], and independent results from a co-seismic stress rotation analysis, which concluded a <1% release of the background stress throughout the *M*8 Tokachi event[7].

We further found a distinct region of specifically low *b*-values, east of the Tokachi high-slip zone (Fig. 1b). The characteristic dimension of unusually low *b*-values in an approximately 0.2°×0.2° area (Fig. 1b) is not as large as that in an approximately 1°×1° area, immediately before the Tohoku earthquake (Fig. 1a). For earthquakes in the circle indicated by 2, into which this distinct region was included, there was a pronounced decade-scale decrease in *b*-values as low as 0.5~0.6 (Fig. 2a). The *b*-values were almost the same as those observed immediately before the Tohoku earthquake. Similar to laboratory observations of low-and-decreasing *b*-values that were previously detected as a fault of a few centimeters in length that approached failure[9,19], this was found for natural earthquakes.

The afterslip of the 2003 Tokachi earthquake mainly occurred in areas adjacent to the coseismic high-slip zone[30,37-39]. Around 2005, most of the afterslip almost ended, revealed by a geodetic observation focusing on interplate slips[39]. Based on another geodetic observation focusing on plate coupling[30], the afterslip decelerated during 2005-2008, and then the two (overriding and subducting) plates started to couple around 2008 (for details of the geodetic observation on interplate coupling, see Methods).

The *b*-values that showed strong temporal heterogeneity in the aftermath of the 2003 Tokachi event returned to an increase in 2004~2005 (Fig. 2b). We note that the start of this increase in *b* remarkably coincided with the end of the major afterslip[39]. The *b*-values that were as high as 0.8~0.9 until 2008-2009 and returned to a decrease in 2008-2009 (Fig. 2a), again coincided with deceleration of the afterslip followed by the start of plate coupling[30]. Stresses in the afterslip area resulted in an increase in *b*, followed by a constantly high *b*-value ($b$=0.8~0.9). Due to the recovery of plate coupling after around 2008, stresses accumulated and *b*-values decreased over time.

**Comparison with seismic quiescence.** Whereas earlier studies lacked an objective definition of quiescence, later studies developed a variety of measures to quantify it, including the *Z*-value method[40,41], ETAS (Epidemic-type Aftershock Sequence) modeling[42,43], the RTL/RTM method[44], and others for California, Japan, Kamchatka, Aleutian Islands, and Alaska. A previous study[42] showed a quantitative technique to detect relative quiescence in seismicity during aftershock sequences, and demonstrated that the relative quiescence of seismicity has been a good index to prognosticate the occurrence of large earthquakes. It was extended to a wider area, following the proposal of the ETAS model as the quantitative expression of standard and non-standard earthquake sequences[43]. The ETAS model can be used to test whether or not there is a change in seismicity pattern at $T_c$ (change point time) to detect non-standard sequences such as seismic activation and quiescence[45] (for details of the ETAS model, see Methods). A previous study[31] that used the ETAS model analyzed earthquakes ($M \geq 5.7$ and all depths) during Jul. 1965-Jun. 2018 in the polygon shown in the inset of Fig. 3a to show that seismic quiescence started during Oct. 2008 while a transition to normal levels had not yet fully occurred, not at least until 2018[46]. This polygon-shaped region, which was used in a previous study[31], includes the distinctly low-*b*-value region east of the Tokachi high-slip zone (Fig. 1b).

We conducted an ETAS analysis to show that seismic quiescence started around Oct. 2008 and continued until present (as of Jun. 2025), based on a catalog from Jul. 1965 to Jun. 2025. To create the left panel of Fig. 3a, we fitted ETAS (red curve) to seismicity with $M \geq 5.7$ and all depths (black curve) during the interval from Jan. 1, 1965 to $T_c$=Oct. 1, 2008, and extrapolated onward. We used transformed time (right panel of Fig. 3a) converted from ordinary time (left panel of Fig. 3a) in such a way that the transformed sequence follows a Poisson distribution with unit intensity. Consequently, the ETAS curve in the left panel of Fig. 3a shows a straight diagonal line in the right panel of Fig. 3a, distinguishing visually the observation rate (black) from the predicting rate (red). Included in this panel are the red curves starting after the change point. These curves are used to show the significance of deviations in the case where the observed cumulative function deviates outside the parabola of 95% significance[45]. The occurrence rates (black) were significantly lower than the extrapolated rates (red) after $T_c$=Oct. 1, 2008. This significant difference continued until recently (Jun. 2025). We noted that the decade-scale duration of seismic quiescence (Fig. 3a) roughly coincides with that of low-and-decreasing *b*-values (from around 2008 to Jun. 2025) in the circle indicated by 2 (Fig. 2b). To assess the uncertainty of the result, the same analysis was conducted for a slightly different magnitude range ($M \geq 5.5$) in Supplementary Fig. 7. The feature for $M \geq 5.5$ did not vary from that for $M \geq 5.7$.

To support the decision to set $T_c$ at Oct. 1, 2008, $T_c$ was searched during the 1960s-2020s (for details of searching appropriate $T_c$, see Methods). When setting $T_c$ at times before Oct. 1, 2008, it was possible to observe significant seismic quiescence for both magnitude ranges ($M \geq 5.7$ and 5.5), but not when setting $T_c$ at times after Oct. 1, 2008 (Supplementary Fig. 8). There was a clear-cut time instant around Oct. 1, 2008 across which seismicity rate changed, implying that setting $T_c$ to Oct. 1, 2008 was appropriate.

To explore the detailed behavior of seismic quiescence, the study region (inset of Fig. 3a) was divided into two sub-regions (one west of the region offshore of the Nemuro peninsula shown in the inset of the left panel of Fig. 3b, and the other east of it in the inset of the right panel of Fig. 3b), and conducted the same analysis as before. The west sub-region includes the zone with an observation of low-and-decreasing *b*-values while the east sub-region does not. For both sub-regions, seismic quiescence appeared. In the west sub-region, the occurrence rates were below the lower margin of 95% confidence intervals of the extrapolation or around this margin (left panel of Fig. 3b). In the east sub-region (right panel of Fig. 3b), the occurrence rates were around the lower margin, where no earthquakes occurred for about 7 years (Oct. 2008-Jun. 2015).

A fundamental question is how quiescence relates to the presence or absence of slow earthquakes. Slow earthquakes are fault slip phenomena that occur over longer time scales than ordinary earthquakes[47], and include low-frequency earthquakes (LFEs), tectonic tremors, which are swarms of LFEs, very-low-frequency earthquakes (VLFs), and slow slip events (SSEs). A specific example that addressed the question includes a case study on a SSE in the focal area of the 1975 Kurile tsunami earthquake[41]. The spatial pattern of quiescence based on the occurrence rate of $M \geq 5$ earthquakes could be explained qualitatively by the changes in Coulomb stress due to this SSE in the focal area of the 1975 tsunami earthquake.

**Comparison with slow earthquake activity.** As was previously suggested[32], the slow earthquake distribution was complementary to co- and post-seismic ruptures of the Tohoku earthquake (for details of slow earthquake data, see Methods and Supplementary Fig. 4). The Japan Trench off the coast of the Tohoku district was divided into three along-strike segments by different behaviors of slow earthquake activity. The co- and post-seismic ruptures of the Tohoku earthquake, which nucleated in the central segment that was characterized by relatively less slow-earthquake activity, were terminated by the two adjacent segments that were characterized by relatively active slow-earthquakes. Inferred from the GNSS observations[48] prior to the Tohoku earthquake, these two segments were only partially coupled, while the central segment was fully coupled although it was eroded by afterslips of the 2003 *M*6.8 and 2008 *M*6.9 Fukushima earthquakes, which suggests that the co-seismic rupture area of the Tohoku earthquake experienced an afterslip of these interplate earthquakes.

We compared the distribution of slow earthquake activity[32] with the high-slip zone of the 2003 Tokachi event[5] (for details of slow earthquake data, see Methods). The result shows that the Tokachi high-slip zone was adjacent to, but hardly overlapped with, the sites where slow earthquake activity occurred (Fig. 4a). To obtain this result, we followed the advice of a previous study[32] and selected tremors and VLFs from slow earthquakes (Supplementary Fig. 4). We also used other indicators of an aseismic slip that were based on observations of repetitive ruptures on the same fault patches, called repeating earthquakes (repeaters)[49] and increases in the seismicity rate without a distinguishable mainshock, called earthquake swarms[50]. We used earthquake swarms containing repeaters, since these are potentially indicative of a spontaneous aseismic slip such as SSEs and afterslips[32,50].

We provide a similar finding from a well-studied tectonic regime such as California. Tremors and LFEs associated with SSEs on a fault segment deeper than the seismogenic zone, where ordinary earthquakes generally occurred, have been observed in the vicinity of the Parkfield section of the San Andreas Fault. On the other hand, the SSEs responsible for the tremors and LFEs have not been reported at Parkfield[51], near which the hypocenter of the 2004 Parkfield earthquake[25] was located. This finding is similar to that for the 2003 Tokachi earthquake.

A comparison between slow earthquake distribution and co-seismic slip areas (asperities) of $M\geq7$ interplate earthquakes[52] showed that slow earthquake distribution is complementary to the asperity distribution, especially in the northern segment[32]. Our observation of the Tokachi events supports the observation of a previous study[32] on the Tohoku and $M\geq7$ earthquakes, which showed that areas hosting frequent slow earthquakes impeded earthquake rupture propagation.

We found that the low-and-decreasing *b*-value region, next to the areas of a large slip during the 2003 Tokachi event did not overlap with the sites of slow earthquake activity (Fig. 4a). This region has a feature similar to slip areas of the 2011 Tohoku and 2003 Tokachi events and asperity areas of $M\geq7$ events.

**Comparison with magnitude-8~9-class earthquakes.** An area of unusually low *b*-values offshore Hokkaido was located between the slip distributions of magnitude-8-class earthquakes (Fig. 4b and Supplementary Fig. 4): the 1973 Nemuro earthquake of moment magnitude $M_w7.8$[34] and the 2003 $M_w8.0$ Tokachi earthquake[5], where $M$ (JMA magnitude) is 7.4 and 8.0 for the former and latter earthquakes, respectively. We wrote $M_w$ and $M$ for the respective magnitude-8-class earthquakes, because $M_w$ is a quantity proportional to the product of the area of the fault and the amount of fault slip, and its physical meaning is clear. We also found that this low-*b*-value area fell in the margin of a far-offshore asperity of another magnitude-8-class earthquake (Fig. 4b), the 1952 $M$8.2 ($M_w$8.1) Tokachi earthquake[33]. To define the asperities of this earthquake, we followed a previous study[52], specifically defining asperity as an area within the value of half the maximum slip. A comparison with the past magnitude-8-class earthquakes shows that the area of low-and-decreasing *b*-values was located in and around the previously-identified longstanding seismic gap. The seismic gap hypothesis states that fault regions where no large earthquakes have recently occurred, are more prone than others to hosting the next large earthquake[53].

Tsunami deposits due to prehistoric tsunamis were found far inland from the coast of Hokkaido where tsunamis generated by the above historical magnitude-8-class earthquakes did not reach. The latest event, which left those tsunami deposits, occurred in the 17th century[13,54-56] (for details of the 17th-century earthquake, see Methods). A previous study[14] estimated the fault model of the 17th-century earthquake (Fig. 4b), explaining those deposits more comprehensively than another study prior to it[13]. The calculated value for $M_w$ was 8.8, where this magnitude-9-class megaquake, which occurred before the start of JMA's observation, has no $M$ (JMA magnitude). This value was roughly consistent with a previous estimate ($M_w$8.5)[13]. Similar to the 2011 Tohoku earthquake, the 17th-century earthquake had the characteristics of a very large amount of slip at the shallow part of the plate interface near the trench[14]. The low-*b*-value zone located within the fault model of the 17th-century earthquake (Fig. 4b) likely fell near or in the margin of a region with a very large amount of slip.

## Discussion

In the zone of low *b*-values, with its characteristic dimension in an approximately 0.2°×0.2° area seen in most recent times (Fig. 1b), the *b*-values since 2008 have shown a gradual decrease over time, to values around 0.5~0.6 (Fig. 2b). This zone fell in the region of seismic quiescence starting in 2008 (Fig. 3). Decreasing *b*-values and seismic quiescence were synchronized in space and time off Hokkaido. Although both phenomena were observed using seismicity data, the magnitude ranges considered are different from each other: $M \geq 2.5$ seismicity was used to observe the former phenomenon while $M \geq 5.7$ seismicity was used to appreciate the latter one. In particular, 0.9% of $M \geq 2.5$ seismicity are $M \geq 5.7$ events during the 2000~2025 period in the circle 2 (14 $M \geq 5.7$ events out of 1641 $M \geq 2.5$ earthquakes in the bottom panel of Supplementary Fig. 6b). Thus, the seismic quiescence (Fig. 3) and the decreasing trend in *b* (Fig. 2b) are based on independent observations. We searched for $M \geq 2.5$ seismicity since 2000 in the western subregion shown in the inset of the left panel of Fig. 3b, conducting the same ETAS analysis with $T_c$=Oct. 1, 2008 as was done for $M \geq 5.7$ seismicity (Fig. 3). For this analysis, we did not consider the eastern subregion shown in the inset of the right panel of Fig. 3b. This is because in this region, at the periphery and outside of the JMA monitoring area (Supplementary Fig. 1), the earthquake detection capacity is low. The observed rate of seismic activity after $T_c$=Oct. 1, 2008 was low, compared to the extrapolated rate from seismic activity before it (Supplementary Fig. 9), showing relative seismic quiescence. The same feature was observed for $M \geq 2.8$ seismicity.

Our results are also synchronized with the spatiotemporal change in interplate coupling based on a geodetic observation[30] (for details of the geodetic observation on interplate coupling, see Methods). This investigation observed previously-reported transient changes in plate coupling. One of the observed changes was that the plate coupling in a subduction-zone section around the 2003-Tokachi-quake high-slip zone was weak until around 2006, due to an afterslip of this event[37,38]. Based on the result of the spatiotemporal change in the interplate coupling obtained in a previous study[30], the plate coupling started to recover after around 2008. Specifically, we pointed out that the coupling after it in the subduction-zone section from the 2003 Tokachi high-slip zone to the area offshore of the Nemuro peninsula, roughly corresponding to the region shown in the inset of Fig. 2b, was equivalent to or even stronger than the coupling before this major event. The feature remains similar, using the most recent geodetic data (until Feb. 2024)[57].

We further pointed out that the low-*b*-value zone was located within the fault model of the magnitude-9-class earthquake in the 17th century[14] (Fig. 4b), and likely near or in the margin of the asperity area of this earthquake. About 400 years have passed since the last event, although the 17th-century-type megaquake occurred with a return period of about 340~380 years, a value obtained based on deposits of 18 tsunamis over about 6,500 years (see Methods for details of the 17th-century earthquake)[15].

We found a large and persistent low-*b*-value structure between the slip distributions of the 2003 Tokachi and 2011 Tohoku events (Fig. 4b). The slip distribution of another magnitude-8-class earthquake, namely the 1968 *M*7.9 ($M_w$8.3) Tokachi earthquake, overlapped with this structure. The two asperities (high-slip areas) of the 1968 Tokachi event[52] are not in a distinct region of specifically low *b*-values. The south asperity hosted two events of the magnitude-7.5 class, the 1931 *M*7.3 ($M_w$7.3) and 1994 *M*7.7 ($M_w$7.7) quakes[52], while the north asperity hosted no events of this class. An average interval of the three sequential events in 1931, 1968, and 1994 for the south asperity was 31.5 years. About 30 years have passed since the magnitude-7.5-class 1994 event occurred for this asperity, and about 56 years have passed since the 1968 magnitude-8-class Tokachi event occurred between the slip distributions of the 2003 Tokachi and 2011 Tohoku events. Assuming a stress-and-*b* correlation[17], our finding of the large and persistent low-*b*-value structure suggests some levels of stresses between low stresses in the high-slip zone of the Tohoku event and high stresses in east of the Tokachi high-slip zone (Fig. 1b).

In this paper, we studied the *b*-values of earthquakes in the subduction zone in northeastern Japan, providing the most comprehensive summary of previous studies and a better understanding of the *b*-values associated with megathrust earthquakes. We used high-quality seismicity data produced by JMA and carried out comparison studies of different portions in the same convergent margin. Near or in the margin of the asperity area of the 17$^{th}$-century megaquake that occurred about 400 years ago with a 340~380-year return period, we found the low-and-decreasing *b*-value zone (indicative of high and increasing stresses) and observed phenomena that provide insight into subsequent large earthquakes: decade-scale seismic quiescence, a longstanding seismic gap, strong plate coupling, and the absence of an overlap between slow earthquake activity and the megaquake source areas. Our study strengthens the notion that megaquakes off Hokkaido have a characteristic recurrence interval, rather than a previous proposition[9] stating that megaquakes occur randomly over time. We also found high *b*-values (indicative of low stresses) in the source area of the Tohoku earthquake after its occurrence. This result, which infers that the stress relieved by the earthquake does not recover very rapidly to levels before the earthquake, was again inconsistent with the previous proposition[9]. We clarified that our findings diverged from the previous study[9], and we discussed likely reasons—most notably the much longer time window and the inclusion of multiple segments—which lead to a different interpretation than that inferred from the shorter dataset[9]. We kept the focus on the broader, data-driven conclusions of our study.

## Methods

**Seismicity dataset.** We analyzed *b*-values using the JMA earthquake catalog (see Data availability) along a roughly 1000 km-long stretch of the Japanese Pacific plate, following the three-dimensional geometry of the subducting slab[58,59]. The JMA catalog includes earthquakes since 1923. Seismic networks have gradually modernized over time. To complement its own network, JMA started in 1997 real-time processing of waveform data from many other networks operated by Japanese universities and institutions. Following a previous study[23], we used seismicity since 2000 and truncated the catalog below $M$=2.5 to discard in advance all data that might not be homogeneous. To consider only seismicity on and around the plate boundary, outer-rise seismicity, east of the trench axis, was removed in advance from the earthquakes catalog.

To enhance seismic-station density offshore of the Pacific coast of eastern Japan, JMA started, in Sep. 2020, to add real-time processing of waveform data from S-net (Seafloor observation network for earthquakes and tsunamis along the Japan Trench[60]) (Supplementary Fig. 1). Cross-sectional views show a comparison between $M \geq 2.5$ seismicity before and after the addition of the S-net data (Supplementary Fig. 2). One feature, hypocenters at distances nearer than ~100 km from the coast showing a distribution roughly consistent with the plate interface, was common to seismicity before and after it. At distances from ~100 km to the trench axis, different features were observed: hypocenters were located in the depth range of roughly ±20 km from the plate interface after the addition of the S-net data, while hypocenters before it were distributed over a wide range of depths. This is due to the difference in seismic-station density between before and after the addition of the S-net data, which caused a difference in depth accuracy of earthquakes (Supplementary Fig. 2). At distances beyond the trench axis (not studied in this paper), the feature of distribution of hypocenters at deep depths was common to seismicity before and after the addition of the S-net data.

Our observation that earthquakes at intermediate distances (from ~100 km to the trench axis) before the addition of the S-net data were distributed over a wide range of depths (Supplementary Fig. 2), was interpreted as an indication of the absence of stations off the Pacific coast. We assumed that earthquakes at the intermediate distances before the addition of the S-net data occurred around the plate interface. Such earthquakes were used as data for our analysis. We decided to use seismicity in the depth range of ±20 km from the plate interface at distances nearer than 100 km from the coast and seismicity from all depths at distances from 100 km to the trench axis.

To support our assumption described above, earthquakes in the JMA catalog were compared with those in the catalog of F-net (Full Range Seismograph Network of Japan), compiled by the NIED (National Research Institute for Earth Science and Disaster Resilience)[60] (Supplementary Fig. 3). The F-net catalog is based on centroid moment tensor determination from broadband seismograms. We paired an earthquake in the F-net catalog with that in the JMA catalog, if the epicentral distance between them was less than 30 km and the time difference between them was within 2 seconds, while one-to-multiple cases were ignored. Paired earthquakes were considered as an identical event. Paired events during a representative period before the addition of S-net data (Jan. 1, 2012 to Dec. 31, 2019) were used for cross-sectional views of earthquakes in the JMA and F-net catalogs (Supplementary Fig. 3). Visual inspection shows that at intermediate distances (from ~100 km to the trench axis), hypocenters for the F-net catalog were located in the depth range of roughly ±20 km from the plate interface, while hypocenters for the JMA catalog were not. We support our assumption that earthquakes at the intermediate distances before the addition of the S-net data occurred around the plate interface.

***b*-value analysis.** To estimate *b*-values consistently over space, we employed the HIST-PPM[61] technique. The *b*-values were calculated at vertexes in Delaunay tessellations where a set of vertexes indicates a set of epicenter's coordinates. To obtain the spatial variation of location-dependent *b*-values, we applied an empirical Bayesian smoothing technique, assuming a piecewise-linear function $b(x,y)$, using Delaunay tessellation with the locations of earthquakes, where $x$ and y are longitude and latitude coordinates, respectively. Namely, we used a piecewise-linear function defined by tessellation with triangles, with coefficient values for the function assigned to earthquake locations and additional boundary points. Hence, a function value at any location in space is uniquely defined by the linear interpolation of values at the three nearest points (earthquakes) that determine a Delaunay triangle. A map view (Fig. 1a) that shows the colored *b*-value points on the epicenters of seismicity indicates that a zone of low *b*-values coincided with the location of a large coseismic slip of the Tohoku earthquake. Another map view (Fig. 1b) shows a zone of low *b*-values, east of the Tokachi coseismic slip zone.

We calculated the timeseries of *b*-values in and around the low-*b*-value volumes indicated by 1 and 2 (Fig. 1). To eliminate doubt about the result obtained by using conventional methods such as the MAXC and EMR methods[62,63] with the choice of moving time-windows, we employed a method that uses a state space model and a particle filter to estimate temporal variations in *b*-values[36]. Note that the parameter, which corresponds to the moving time-windows introduced when using the Maximum Curvature (MAXC) and Entire Magnitude Range (EMR) methods and determines the ability to adapt to variation in *b*-value, is automatically adjusted from the data to the optimum value in the method[36]. The MAXC and EMR methods are limited to a constant *b*-value and do not consider its timeseries structure. In contrast, the method of the state space model using the particle filtering assumes the timeseries structure. To create Fig. 2, we used seismicity in and around the low-*b*-value volumes indicated by 1 and 2 in Fig. 1, where the median (curve) and 68% area (vertical segments) of the posterior distribution of the *b*-value estimated by a particle filter were plotted for $M \geq 2.5$ (blue) and 2.8 (red).

Overall, the patterns for $M \geq 2.5$ and 2.8 are similar in both volumes (Fig. 2). An exception occurs in volume 1 (Fig. 2a) during several years immediately following the Tohoku earthquake, where the two magnitude thresholds yield fluctuating and somewhat divergent *b* estimates; thereafter, they converge to similar values ($b \sim 1$). These large post-mainshock fluctuations are consistent with highly active aftershock sequences, a commonly observed behavior. During such intervals, the completeness magnitude can influence the slope of the GR *N*–*M* relation. Future work may therefore benefit from applying the *b*-positive method to obtain estimates less sensitive to completeness when we will focus on *b* values during intense aftershock sequences.

One possible avenue to further strengthen our analysis is to consider different *b*-value estimation approaches, such as likelihood-based approach presented by Kamer and Hiemer[64] that might be feasible for the objective of capturing detailed aspects of *b*-value variation in space and time.

**ETAS model.** The ETAS model[43] is a point-process model that represents the activity of earthquakes of a minimum magnitude ($M_{th}$) and above in a certain region during a specified time interval. Seismic activity includes the background activity at a constant occurrence rate μ (Poisson process). The model assumes that each earthquake (including the aftershock of another earthquake) is followed by aftershocks. Aftershock activity in the time domain is represented by $\lambda=k(c+t)^{-p}$, where $t$ is time, λ is the number of aftershocks with $M \geq M_{th}$ per unit time at $t$, and $p$, $c$, and $k$ are constants. The rate of an aftershock occurrence at $t$ following the $i$-th earthquake (time $t_i$ and magnitude $M_i$) is given by $\nu_i(t)=K_0\exp\{\alpha(M_i-M_{th})\}(t-t_i+c)^{-p}$ for $t>t_i$, where $K_0$ and α are constants. $K_0$, α, $c$, and $p$ are common to each target aftershock sequence in a region. The rate of occurrence of the whole earthquake series at $t$ becomes $\lambda(t|H_t) = \mu + \sum_{t_i<t} \nu_i(t)$. The summation is performed for all $i$ satisfying $t_i<t$. Here, $H_t$ represents the history of occurrence times with associated magnitudes from the data $\{(t_i, M_i)\}$ before time $t$. The parameter set θ=(μ, $K_0$, α, $c$, $p$) represents the characteristics of seismic activity. The units of the parameters are day$^{-1}$, day$^{-1}$, no unit, day, and no unit, respectively. For the case of $K_0$=0, the ETAS model reduces to the Poisson process. The parameter set θ is estimated using the maximum likelihood method. *AIC* (Akaike Information Criterion), useful for assessing the goodness of fit of the model to data, is computed from the maximum log-likelihood and the number of adjusted parameters (5 parameters in this study). Because $K_0$ depends on $M$ in the model, it is necessary to assume a magnitude at which a value for $K_0$ needs to be known. Throughout this study, $M$=9.0 was assumed for estimating $K_0$. For fitting the ETAS model, two time intervals were considered. One interval is called the target interval for which the ETAS model parameters are computed. The seismicity in this period may be affected by earthquakes which occurred before this period due to the long-lived nature of aftershock activity. To consider this effect, the other time interval, which is precursory to the target interval (called the precursory interval) is chosen and aftershock activities following earthquakes in this period are considered in the computation. The target interval is defined by $S$ to $T$, for which the ETAS model parameters are computed. The precursory interval is defined by 0 to $S$.

Using the maximum likelihood estimate, it is possible to visualize how well or poorly the model fits an earthquake sequence by comparing the cumulative number of earthquakes with the rate calculated by the model. Ordinary time can be converted to transformed time in such a way that the transformed sequence follows the Poisson process (uniformly distributed occurrence times) with unit intensity (occurrence rate) so that visualization can be achieved in two ways[45]: one graph using ordinary time and the other using transformed time (Fig. 3). If the model presents a good approximation of observed seismicity, an overlap with each other is expected. Furthermore, a plot of the cumulative number of earthquakes against the transformed time shows a nearly straight line.

We examined that seismicity pattern changed at a particular time (change point time, $T_c$). For this examination, the observed cumulative number of earthquakes was compared with the modeled cumulative number of earthquakes during the prediction interval defined by $T$ to $T_{end}$ ($>T$). Here, $T$ is equal to $T_c$, the prediction interval is adjoint to the target interval (from $S$ to $T$), and θ used for this model is θ estimated for the target interval[45]. If the model presents a good approximation of observed seismicity during the target interval, and if the cumulative number of earthquakes deviates from the rate calculated by the model during the prediction interval, then the seismicity pattern is considered to change at $T_c$. Included in the graph using the transformed time is the parabola of 95% significance. When the number of earthquakes is insufficiently large, significance actually depends on sample size due to the estimation accuracy of the parameters. The 95% significance or higher significance is defined in the case where the empirical curve is around the parabola or deviates outside the parabola.

We used a GUI-based program package XETAS[45]. In Fig. 3 and Supplementary Fig. 7, according to a previous study[31], we set Jul. 1, 1965 00:00 as 0 days. Then, we set $S$=0.01 days (elapsed time since Jul. 1, 1965 00:00) and $T$=$T_c$=15,798 days (corresponding to Oct. 1, 2008 00:00)[31]. Considering the catalog that includes earthquakes until Jun. 23, 2025 00:00, we set $T_{end}$=21,907 days (corresponding to Jun. 23, 2025 00:00).

**Searching appropriate $T_c$.** We supported the choice of $T_c$=15,798 days (corresponding to Oct. 1, 2008)[31] by conducting a test based on *AIC*. We focused on differences between the standard (single) ETAS model and an extended two-stage ETAS model, which covers non-standard cases in fitting seismicity timeseries[45]. The two-stage ETAS model is the simplest alternative to the standard (single) model. This two-stage model assumes different θ values in subperiods before and after $T_c$, and the whole period is divided into two adjoining periods to fit the ETAS models separately. Testing whether or not there are changes in seismicity pattern at $T_c$ in a given period is deduced to the problem of model selection, using *AIC*. We compared $AIC_{single}$ (*AIC* for a standard (single) ETAS fitting) with $AIC_{2stage}$ (*AIC* for a two-stage ETAS fitting) to select the model with the smaller value, where $AIC_{2stage}=AIC_1+AIC_2$ ($AIC_1$ and $AIC_2$ for fitting to the subperiod before $T_c$ and the subperiod after $T_c$, respectively). If $T_c$ is searched from the data, the two-stage model becomes harder to accept. Namely, $AIC_{single} \geq AIC_{2stage}+2q$ implies that the two-stage model is selected, where $q$ is the degree of freedom to search for the best candidate $T_c$ from the data. $q$ depends on sample size (number of earthquakes in the target period)[65]: $q$ increases with sample size and, for example, lies in the 4-5 range for sample size between 100 and 1000. $T_c$ with $AIC_{single}<AIC_{2stage}+2q$ cannot be considered as a candidate for the instant of seismicity change.

$\Delta AIC$ (=$AIC_{single}$-$AIC_{2stage}$), as a function of $T_c$ during the 1960s-2020s (Supplementary Fig. 8), shows $\Delta AIC$-values around $2q$ or above from mid-1970s to $T_c$=15,798 days (Oct. 1, 2008, indicated by a black vertical line) for $M \geq 5.5$ ($M_{th}$=5.5) in the blue curve. To create this figure, seismicity in the region in the inset of Fig. 3a was used. Similarly, we observed several $\Delta AIC$-values around $2q$ or above from the mid-2000s to $T_c$=15,798 days for $M \geq 5.7$ ($M_{th}$=5.7) in the red curve. Given that $\Delta AIC$ was below $2q$ after $T_c$=15,798 days for both magnitude ranges, times around 15,798 days were considered as the most recent times for $T_c$. There was a discontinuation in smoothing around Oct. 1, 2008, across which seismicity rate changed, supporting that the choice of $T_c$=Oct. 1, 2008 in Fig. 3 and Supplementary Fig. 7 was appropriate. This was not the case for $M \geq 6.0$ ($M_{th}$=6.0) in the green curve, probably due to low seismicity in our case. $\Delta AIC$ for $M \geq 6.0$ also showed a better (but insignificant) outcome when the two-stage ETAS model, rather than the single ETAS model, was used (Supplementary Fig. 8). Namely, $\Delta AIC$ was higher than 0 ($\Delta AIC > 0$) when $T_c$ was around 15,798 days and before, but it was below the horizontal dashed line indicating $2q$. This line indicates a hurdle to the selection of the two-stage ETAS model, given that $T_c$ was searched from the data during 1960s-2020s.

It was difficult to stably obtain $AIC_{2stage}$-values for all times during the 1960s-2020s where seismicity in the region in the inset of Fig. 3a was considered. To reduce the number of optimized parameters $\theta = (\mu, K_0, \alpha, c, p)$, the $p$-value obtained for the single ETAS fitting was assumed to be applicable for the two-stage ETAS fitting. Namely, the $p$-value, which was independent of $T_c$, was common to the two subperiods considered to compute $AIC_1$ and $AIC_2$. Given the predefined $p$=1.060 for $M_{th}$=5.5, 1.041 for $M_{th}$=5.7, and 1.032 for $M_{th}$=6.0, the set of parameters ($\mu$, $K_0$, $\alpha$, $c$) were optimized for each $T_c$ and each $M_{th}$ to compute $AIC_1$ and $AIC_2$, resulting in $AIC_{2stage}$ ($=AIC_1+AIC_2$) (Supplementary Fig. 8).

**Slow earthquake data.** Because slow earthquakes occurred in the vicinity of megathrust earthquakes, shear deformation of these different earthquakes interact[32]. Slow earthquakes sometimes preceded large earthquakes and were related to the rupture termination of large earthquakes. Moreover, areas hosting frequent SSEs impeded the rupture propagation of large earthquakes. Elucidating the entire spectrum may help us to infer the occurrence time and rupture extent of large earthquakes[32].

We compared the distribution of areas of low $b$-values and distributions of slow earthquakes (tectonic tremors, VLFs, earthquake swarms) in Fig. 4a (see Supplementary Fig. 4a). We used three slow-earthquake catalogs provided by Nishikawa et al.[32]: the tectonic-tremor catalog, the VLF catalog, and the earthquake-swarm catalog.

These authors detected tectonic tremors, using the envelope correlation method and using continuous seismograms from 150 S-net stations (Supplementary Fig. 1) that covered the Aug. 2016 to Aug. 2018 time period. Because the results included many ordinary earthquakes, procedures to reduce false detection of ordinary earthquakes were implemented. The resultant tectonic-tremor catalog included events with a duration of 20-300 s. Following Fig. 3 of Nishikawa et al.[32], tremors with a duration of 80 s or longer were used for this paper.

Nishikawa et al.[32] detected VLFs using the matched-filter-based method published in Matsuzawa et al.[66] and applied this method to the F-net broadband seismograms from Jan. 2014 to Aug. 2018. Nishikawa et al.[32] obtained the VLF catalog that includes VLFs during the period from Jan. 2005 to Aug. 2018 by merging the result obtained by these authors[32] with the result from Jan. 2005 to Dec. 2013 in Matsuzawa et al.[66].

Nishikawa et al.[32] followed the method of a previous study[50] for earthquake swarm detection and used $M \geq 3$ events in the JMA catalog. Then, these authors[32] created an earthquake-swarm catalog during the periods from Jan. 1991 to Dec. 2010 and from Jan. 2014 to Aug. 2018, where extremely intensive aftershocks of the 2011 Tohoku earthquake prevented them from analyzing seismicity from 2011 to 2013. To create the catalog, first, seismic sequences with seismicity rates that are inconsistent with the space-time ETAS model were classified as earthquake swarms; second, earthquake swarms containing at least one repeater were extracted. The repeaters, which are repetitive rupture sequences on the same fault patch and are often used as a creep meter on the plate interface, were observed to accompany geodetically detected slow slip events in subduction zones. Nishikawa et al.[32] focused on earthquake swarms containing repeaters, and ensured the occurrence of aseismic slip transients during the time periods of the earthquake swarm activities.

The repeaters described in the last paragraph were detected based on the waveform similarity method of Uchida and Matsuzawa[67], applying this method to earthquake waveforms observed by the micro-earthquake observation network of several universities for the 1984 to Aug. 2016 time period. Nishikawa et al.[32] obtained the repeater catalog by merging events detected by themselves and those detected by Uchida and Matsuzawa[67].

**17th-century earthquake.** The fault model of the 17th-century $M_w$8.8 earthquake in Fig. 4b (see Supplementary Fig. 4) was obtained from Ioki and Tanioka[14]. Those authors estimated the fault model to explain the lowland widespread tsunami deposit data and tsunami deposit data at a high cliff near the coast. A previous study[56] estimated the fault model of the same earthquake, which reproduced only the former data, but not the latter one. $M_w$ is comparable between the two studies[14,56], but $M_w$8.8 obtained by Ioki and Tanioka[14] is slightly larger than the $M_w$8.5 obtained in the previous study[56]. The characteristic of the fault model[14] indicates that the corresponding earthquake ruptured a large area off Hokkaido and a very large slip amount (at least 25 m) at the shallow part of the plate interface near the trench.

Previous studies found prehistoric tsunamis. One such study[54] found deposits by nine tsunamis over about 4,000 years, while another study[68] found deposits by 15 tsunamis over about 6,000 years. The recurrence interval of those events is about 400-600 years[69]. The latest event occurred in the early 17th century because the latest tsunami deposits were found just under the volcanic ash caused by the 1677 Tarumae eruption[54].

The Headquarters of Earthquake Research Promotion found deposits by 18 tsunamis over about 6,500 years[15]. Using a long-term evaluation for a 17th-century-type earthquake, the average recurrence interval was evaluated at about 340-380 years[15]. Since about 400 years have passed since the 17th-century earthquake, it is likely that the occurrence of a 17th-century-type earthquake is imminent[15].

**Geodetic observations on interplate coupling.** Iinuma[30] proposed a monitoring method to grasp spatiotemporal changes in interplate coupling in a subduction zone based on the spatial gradients of surface displacement rate field. Those spatiotemporal changes in interplate coupling were estimated along the plate boundary in northeastern Japan by applying the proposed method to surface displacement rates based on global positioning system observations during 1997-2016[30]. Iinuma pointed out, based on Fig. 10 of his paper, transient changes in interplate coupling that had been reported in previous studies: (i) weak coupling at the plate interfaces before the Tohoku earthquake offshore Fukushima (south of a large slip area of this earthquake), (ii) recovery process of coupling at and around the rupture area of the 1994 Sanriku-Haruka-oki earthquake, and (iii) weak plate coupling around 2006 corresponding to an afterslip of the 2003 Tokachi earthquake[37,38]. We note, from the same figure, that interplate coupling offshore Hokkaido after the period of this afterslip, namely on a subduction-zone section from offshore of the Nemuro peninsula to the high-slip zone of the 2003 Tokachi earthquake after 2008, was equal to or even stronger than that before the occurrence of the same earthquake. A feature similar to this remains observable in recent times, based the results obtained by the Geospatial Information Authority of Japan[57], which used the same method as that proposed by Iinuma[30], applying it to data until Feb. 2024, although a section east of the offshore Nemuro peninsula was not included in that study's regions.

**Data availability.** The JMA earthquake catalog is available at https://www.mri-jma.go.jp/Dep/sei/fhirose/plate/en.index.html. The F-net catalog is available at https://www.fnet.bosai.go.jp/top.php?LANG=en. Depths of the upper boundary of the Pacific Plate were obtained from https://www.mri-jma.go.jp/Dep/sei/fhirose/plate/en.index.html[58,59]. Data of slow earthquake activity were obtained from https://www.science.org/doi/10.1126/science.aax5618[32]. The authors digitized co- and post-seismic slip distributions (2011 Tohoku earthquake[35]; 2003 Tokachi earthquake[5]), outlines of coseismic slip areas (1973 Nemuro earthquake[34]; 1968 Tokachi earthquake[52]; 1952 Tokachi earthquake[33]), outlines of asperity areas (1973 Nemuro earthquake[34]; 1952 Tokachi earthquake[33]), and an outline of the fault model of the 17th-century earthquake[14]. These digitized data were deposited to Figshare at https://doi.org/10.6084/m9.figshare.30657251.v3. An outline of asperity areas of the 1968 Tokachi earthquake[52] was obtained from https://www.aob.gp.tohoku.ac.jp/~uchida/page_3_asp-e.html. Seismic station data used for Supplementary Fig. 1 were obtained from https://www.jishin.go.jp/database/observation_station/spots_2022/ and https://www.jishin.go.jp/main/kansoku/kansoku13/index.htm. Source data for Figs. 1-4 were deposited to Figshare at https://doi.org/10.6084/m9.figshare.30569630.v3.

**Code availability.** The software package HIST-PPM[61], used for Figs. 1 and 4 and Supplementary Fig. 5, was obtained from https://www.ism.ac.jp/editsec/csm/index_j.html. The source code of the method that uses a state space model and a particle filter, used for Fig. 2 and Supplementary Fig. 6, is available at the GitHub repository (https://github.com/D-I-29/gr-b-pf). The Generic Mapping Tools (GMT), used for Figs. 1, 3, and 4 and Supplementary Figs. 1-5, 7, and 9, are an open-source collection (https://www.generic-mapping-tools.org). The seismicity analysis software package ZMAP[70], used for the lower right insets of Fig. 1, were obtained from https://github.com/CelsoReyes/zmap7.

**Acknowledgments**

This study was partially supported by the Ministry of Education, Culture, Sports, Science and Technology (MEXT) of Japan, under The Second Earthquake and Volcano Hazards Observation and Research Program (Earthquake and Volcano Hazard Reduction Research) (K.N., T.H.) and under STAR-E (Seismology TowArd Research innovation with data of Earthquake) Program Grant Number JPJ010217 (K.N., T.H.). We thank Y. Noda for help with implementing HIST-PPM.

**Author Contributions:** K.Z.N. and T.H. designed the research. K.Z.N. acquired data. K.Z.N. and D.I. analyzed data. K.Z.N., T.H., and D.I. contributed to writing the paper.

**Competing interests:** The authors declare no competing interests.

## Additional information

**Supplementary Information:** The online version contains supplementary material available at ######.

**Correspondence** and requests for materials should be addressed to K. Z. Nanjo.

**Peer review information:** Communications Earth & Environment thanks four reviewers for their contribution to the peer review of this work. A peer review file is available.

## Figure captions

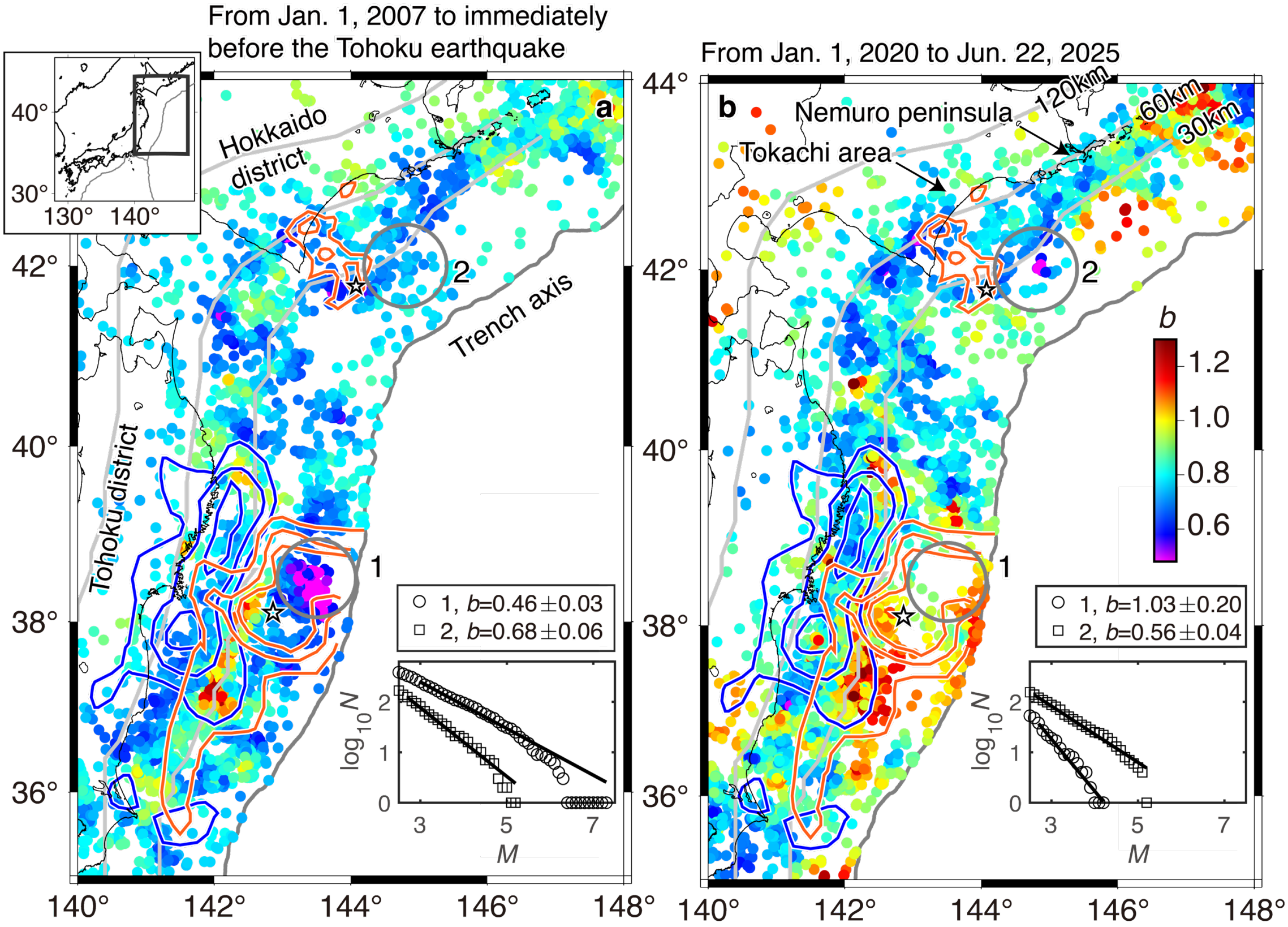

**Fig. 1. *b*-value maps.** (**a**) Map of *b*-values computed at events with $M \geq 2.8$ during the period from 2007 to immediately before the 2011 *M*9.0 Tohoku earthquake. Stars off Tohoku and Hokkaido districts indicate the epicenters of the Tohoku earthquake and the 2003 *M*8.0 Tokachi earthquake, respectively. Orange contour lines indicate coseismic slips of these earthquakes[5,35]. Blue contour lines indicate post-seismic slips of the Tohoku earthquake[35]. The circles with a radius of $r$=50 km, indicated by 1 and 2, show the regions considered in Fig. 2. Dark grey curve indicates the trench axis and light grey curves indicate depth contours of the plate boundary[58,59]. Upper left inset shows the study region (black rectangle). Lower right inset shows the *N*-*M* distribution of earthquakes in the circles indicated by 1 and 2. (**b**) Same as **a** for seismicity from 2020 to the present. Nemuro peninsula and Tokachi area are indicated by arrows.

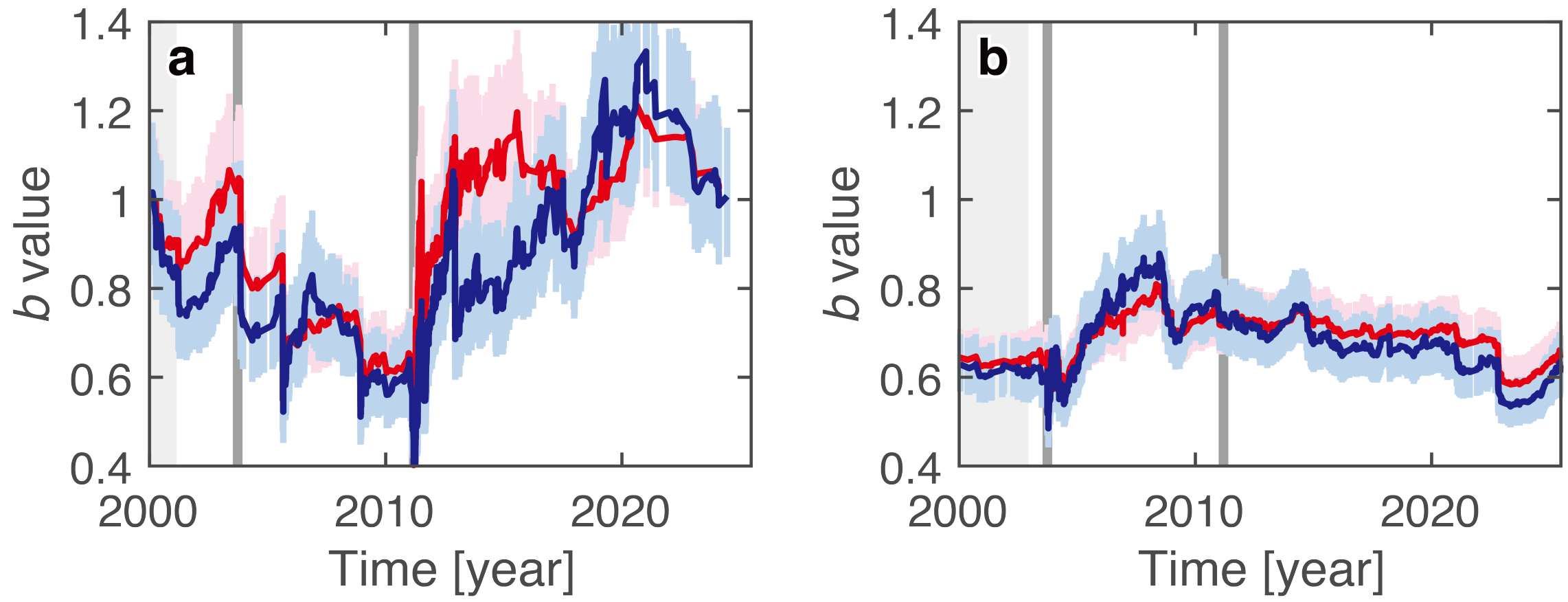

**Fig. 2. *b*-value timeseries.** (**a**) Plot of $b$ as a function of time, as obtained from seismicity data of $M \geq 2.5$ (blue) and 2.8 (red) for the period from 2000 until the present (Jun. 22, 2025) falling in the circle ($r$=50 km) indicated by 1 in Fig. 1. Estimation of $b$-values based on less than 30 earthquakes is unstable so that $b$-values in the grey region, which indicates the period where the first 30 earthquakes occurred, are less reliable. For details of $b$-value analysis, see Methods. Thick vertical lines indicate the moments of the 2003 Tokachi and 2011 Tohoku earthquakes. (**b**) Same as **a** for the circle indicated by 2 in Fig. 1.

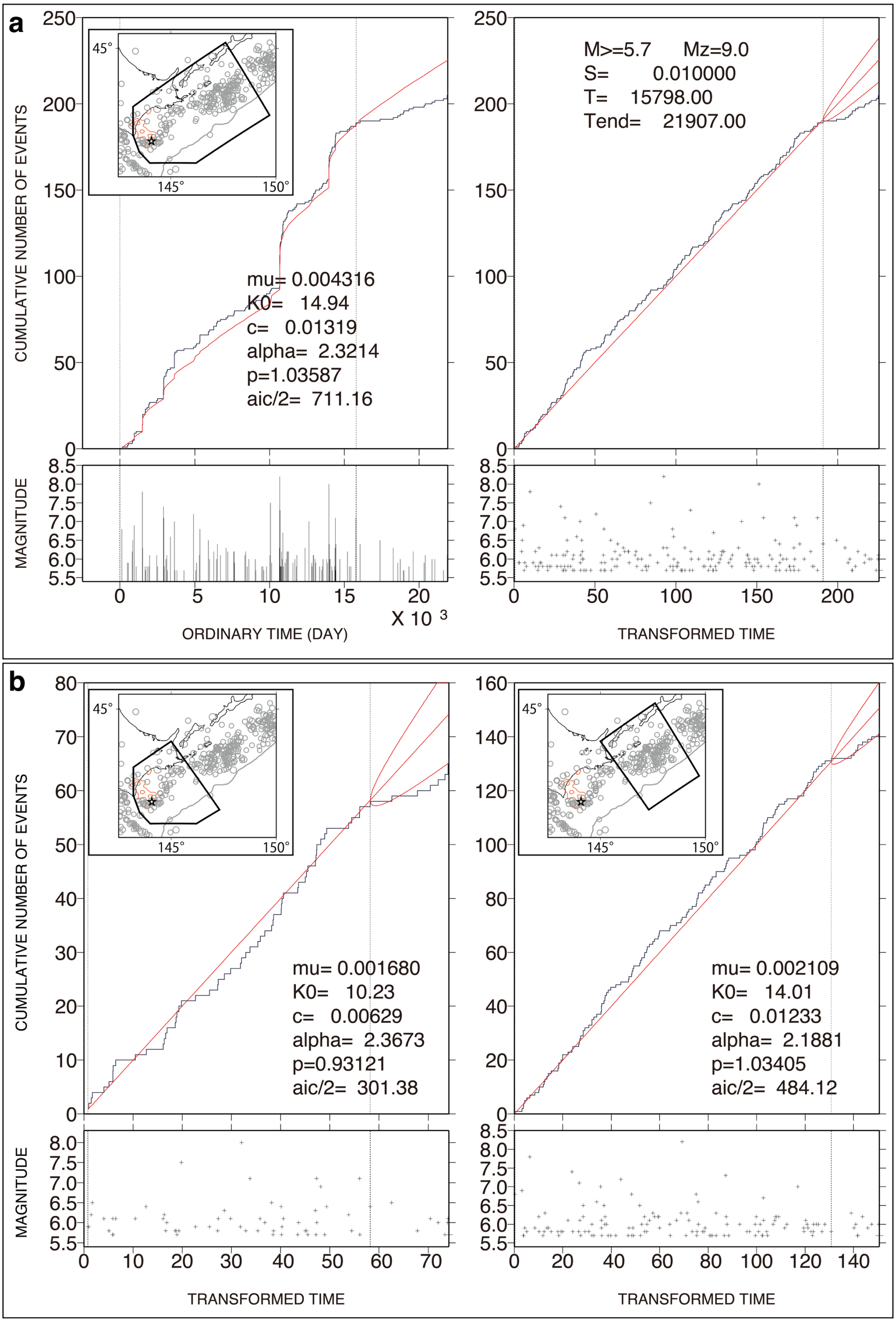

**Fig. 3. Changes in seismicity rate.** (**a**) Cumulative function of $M \geq 5.7$ earthquakes is plotted against ordinary time (left panel) and transformed time (right panel), showing the ETAS fitting in the target interval from Jul. 1, 1965 until Oct. 1, 2008 and then extrapolated until Jun. 22, 2025. The parabola represents the 95% confidence intervals of the extrapolation (for details of the ETAS model, see Methods). The smaller panel below each larger panel indicates an *M*-time diagram. Inset indicates $M \geq 5.7$ earthquakes since Jul. 1965 and the study region[31]. Also included are the epicenter and slip contour lines of the 2003 Tokachi earthquake[5]. (**b**) Left and right panels: same as the right panel of **a** except that the study regions in the left and right panels are the regions west and east of the Nemuro peninsula coast, respectively.

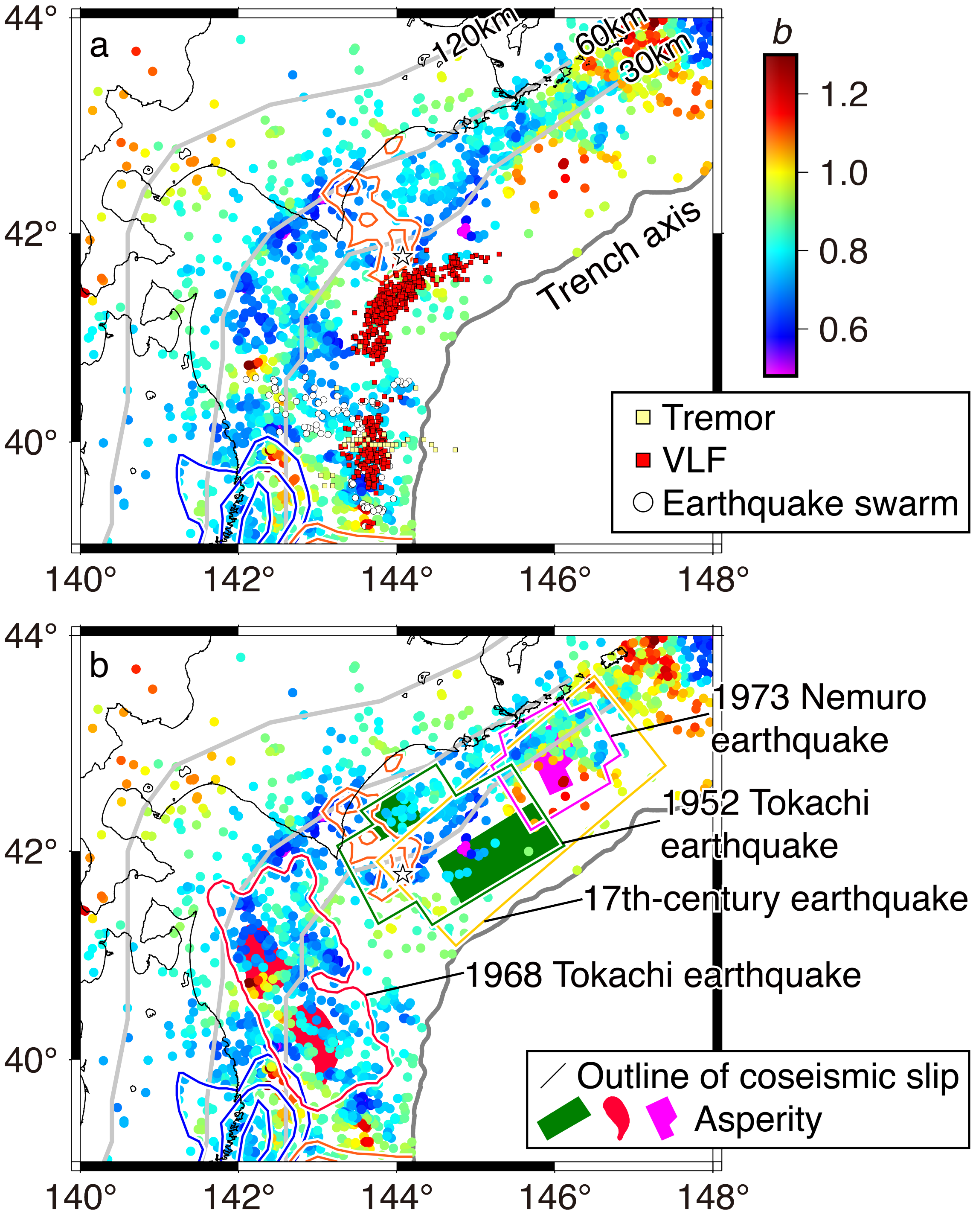

**Fig. 4. Comparison of *b*-values with slow earthquakes and magnitude-8~9-class earthquakes.** The *b*-value map is the same as that shown in Fig. 1**b**. For explanations of the star, orange contours, blue contours, and dark and light grey curves, see the caption of Fig. 1. (**a**) Yellow squares, red squares, and white circles indicate epicenters of the tectonic tremors, VLFs, and earthquake swarms containing repeaters, respectively[32]. (**b**)

Thin lines show outlines of the coseismic slip of the magnitude-8~9-class earthquakes: 1973 $M7.4$ ($M_w7.8$) Nemuro earthquake[34] in pink, 1968 $M7.9$ ($M_w8.3$) Tokachi earthquake[52] in red, 1952 $M8.2$ ($M_w8.1$) Tokachi earthquake[33] in green, and 17th-century earthquake of $M_w8.8$[14] in yellow, where $M$ is the JMA magnitude, and $M_w$ is the moment magnitude. Asperity is indicated by a filled area, where asperity is defined as the area within the value of half the maximum slip of an $M_w \geq 7$ interplate earthquake[52]. Also see Supplementary Fig. 4 for slow earthquakes and magnitude-8~9-class earthquakes.

Supplementary Information

# Non-randomness of Japan megaquakes implied by stress recovery and accumulation

Kazuyoshi Z. Nanjo, Takane Hori, and Daichi Iwata

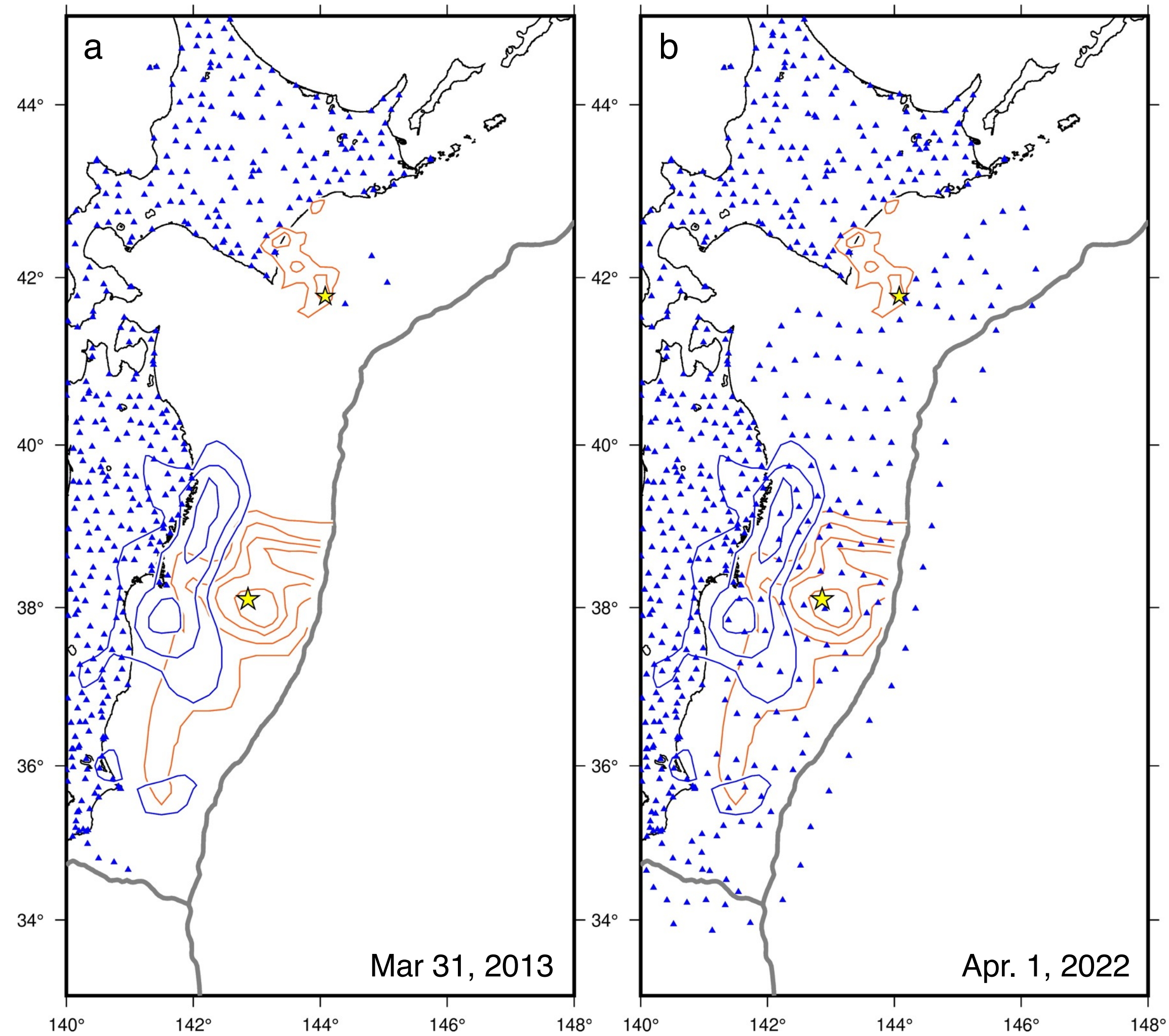

**Supplementary Fig. 1.** Distribution of stations in operation on Mar. 31, 2013 in **a** and Apr. 1, 2022 in **b** (see Data availability). For an explanation of stars, contour lines, and grey curves, see the caption of Fig. 1.

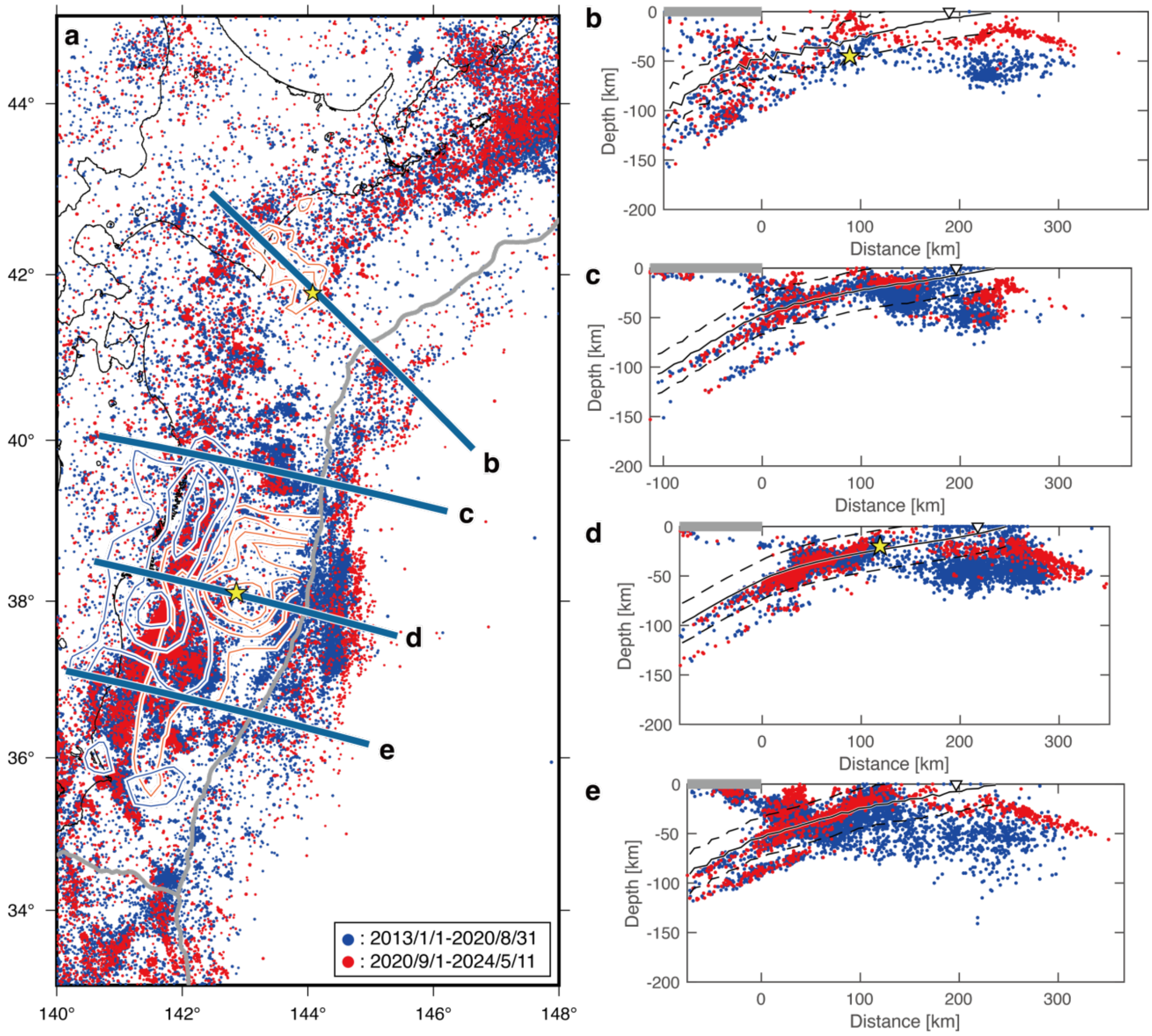

**Supplementary Fig. 2.** Spatial distribution of earthquakes before and after adding the S-net data. (**a**) Red and blue dots indicate $M \geq 2.5$ earthquakes (depth≤200 km) during the periods Jan. 1, 2013-Aug. 31, 2020 and Sep. 1, 2020-May 11, 2024, respectively. Cross-sectional views along the line segments, indicated by **b**-**e**, are shown in **b**-**e**. For an explanation of stars, contour lines, and grey curve, see the caption of Fig. 1. (**b**-**e**) Plot of depths of earthquakes along the line segments, indicated by **b**-**e**, with a width of 100 km as a function of distance from the coast, where the right end of the grey bar at depth of 0 km indicates the coast. Black curve indicates the upper surface of the Pacific plate[58,59]. Dashed curves indicate 20 km above and below this surface. Triangle indicates the trench axis.

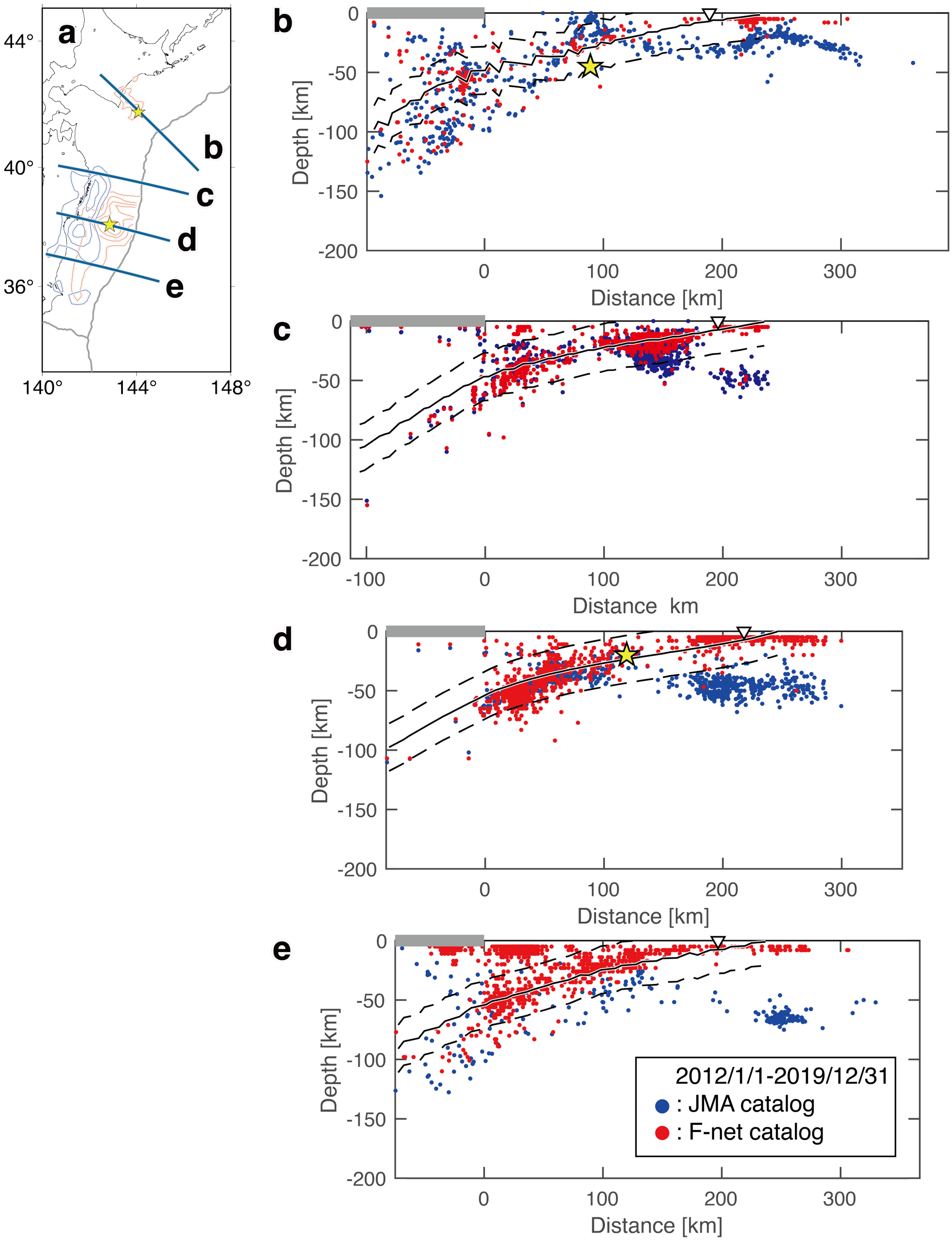

**Supplementary Fig. 3.** Spatial distribution of earthquakes obtained from the F-net catalog (red dots) and the JMA catalog (blue dots). An earthquake in the F-net catalog was paired with that in the JMA catalog if the time difference between them was within 2 seconds and the epicentral distance between them was less than 30 km, while ignoring one-to-multiple cases. Paired earthquakes during the period from Jan. 1, 2012 to Dec. 31,

2019 (depth≤200 km) were used to create the graphs in **b**-**e**. (**a**) Spatial map of line segments, indicated by **b**-**e**. For the explanation of stars, contour lines, and grey curves, see the caption of Fig. 1. (**b**-**e**) Plot of depths of earthquakes along the line segments with a width of 100 km as a function of distance from the coast, where the right end of the grey bar at a depth of 0 km indicates the coast. For an explanation of the black curve, dashed curves, and triangle, see the caption of Supplementary Fig. 2.

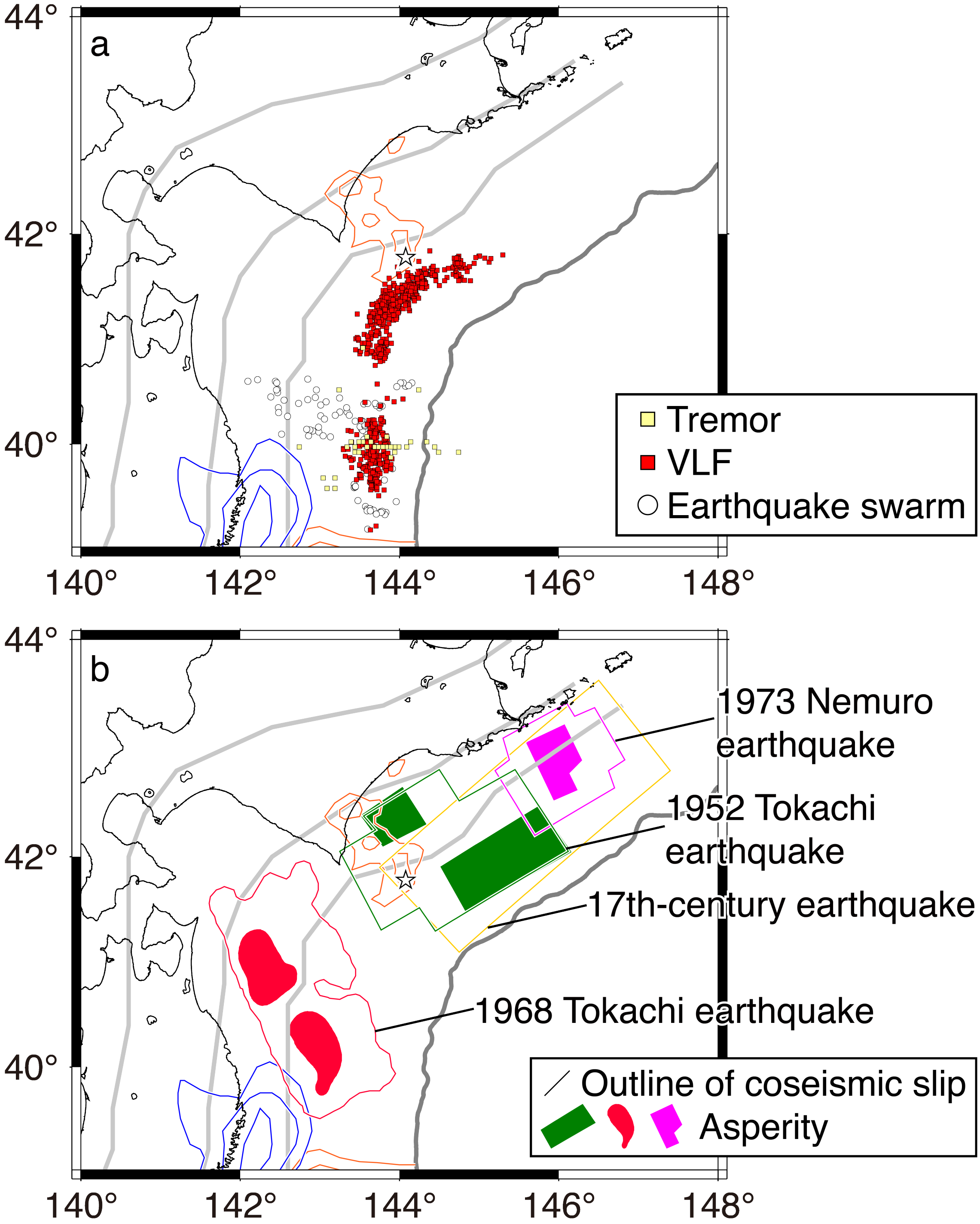

**Supplementary Fig. 4.** Same as Fig. 4 but the *b*-value map was excluded.

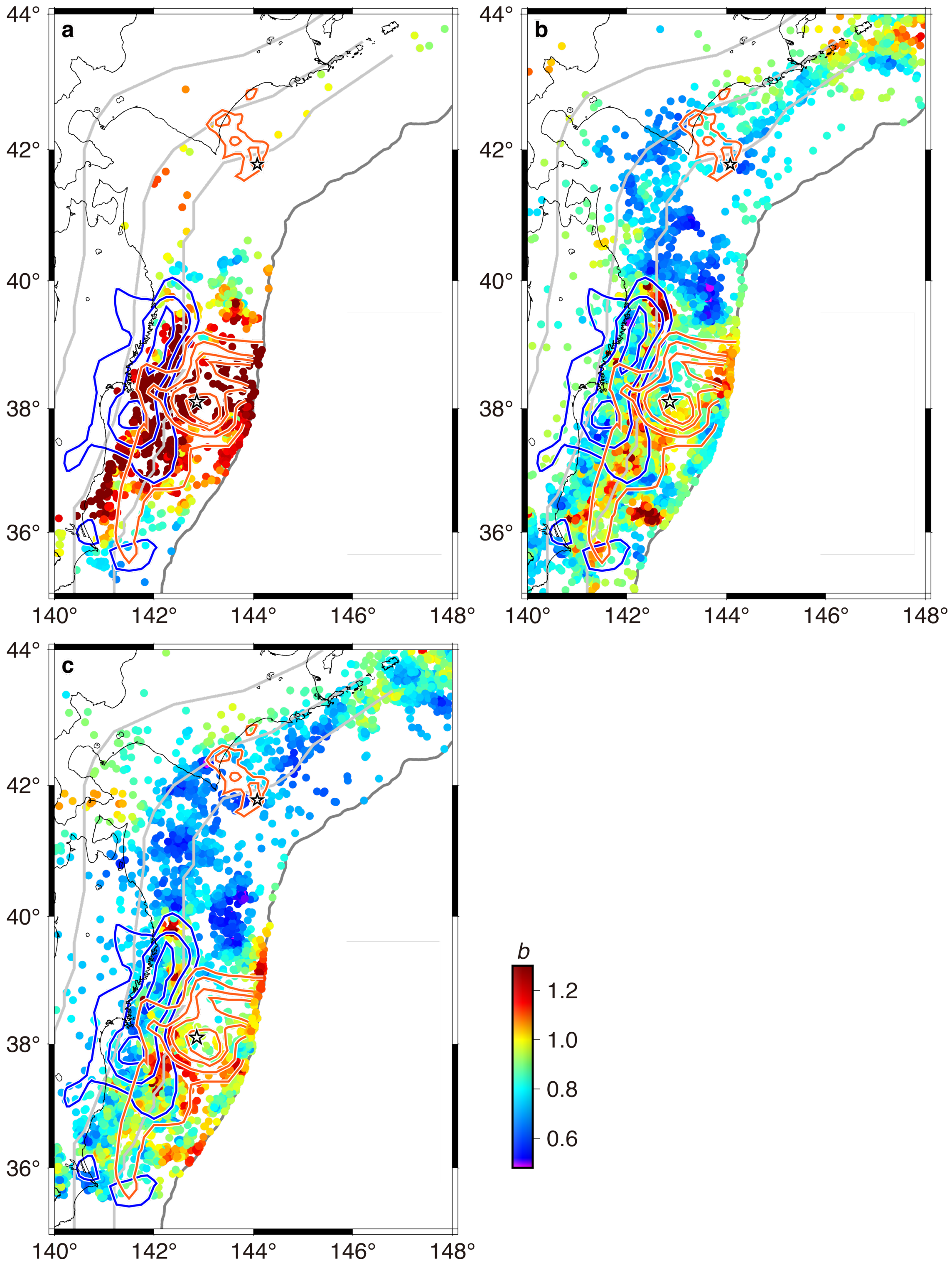

**Supplementary Fig. 5.** Distributions of *b*-values for the three periods: (**a**) from Mar. 26, 2011 to Jun. 30, 2011 ($M \geq 4.0$), (**b**) from Jan. 1, 2012 to Dec. 31, 2015 ($M \geq 3.0$), and (**c**) from Jan. 1, 2016 to Dec. 31, 2019 ($M \geq 2.8$). The period for **a** is the same as that used by a previous study[9] to show 3 months of the highly active Tohoku aftershocks.

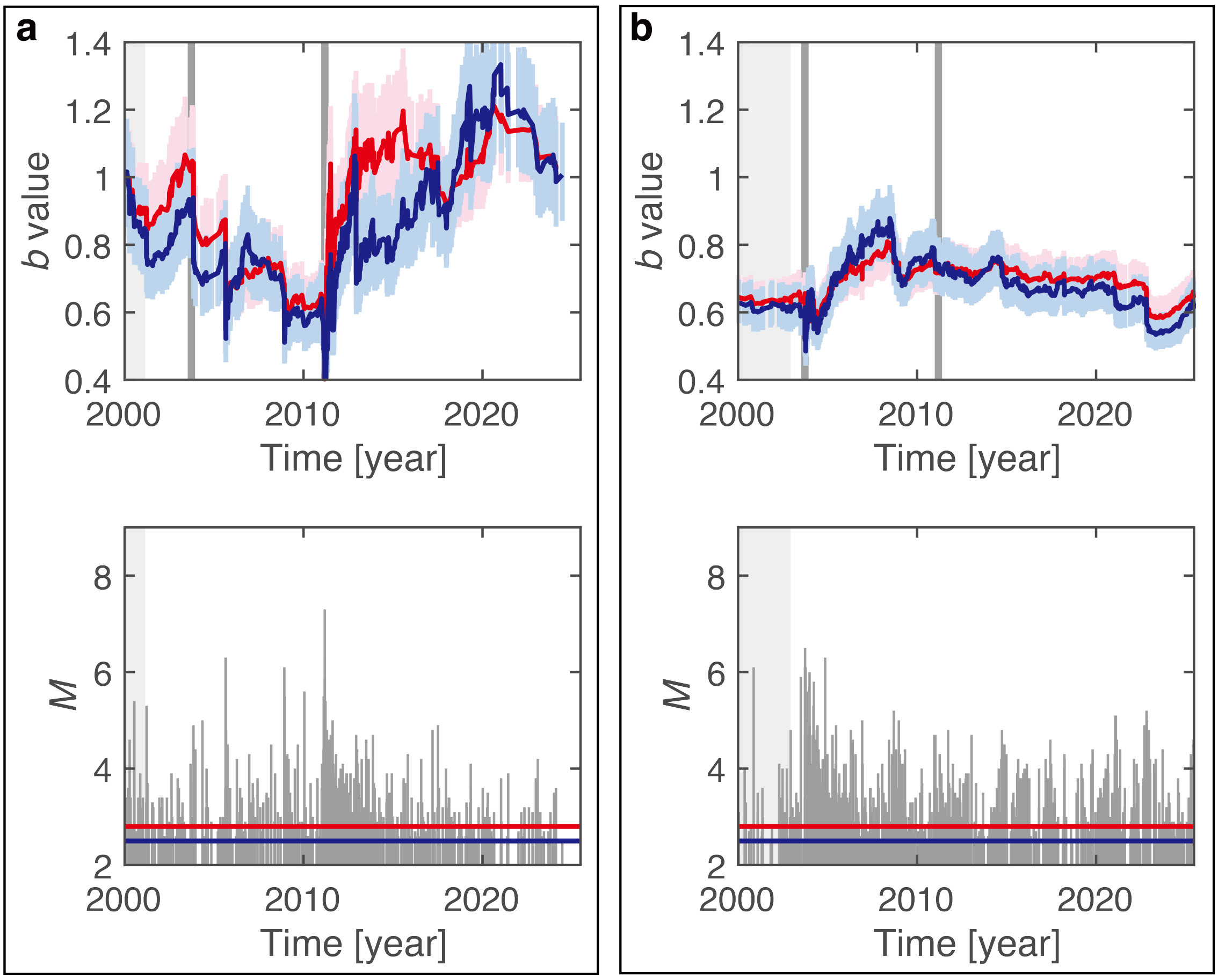

**Supplementary Fig. 6.** Timeseries of *b*-values and *M*-time diagram. (**a**) Top panel is the same as Fig. 2**a**. For explanation of red and blue curves, red and blue vertical segments, light grey region, vertical grey lines, see the caption of Fig. 2 and Methods. Bottom panel is the *M*-time diagram for earthquake data used to create the top panel. Horizontal blue and red lines indicate *M*=2.5 and 2.8, respectively. (**b**) Top panel is the same as Fig. 2**b**. Bottom panel is the same as that in **a** for earthquake data used to create the top panel.

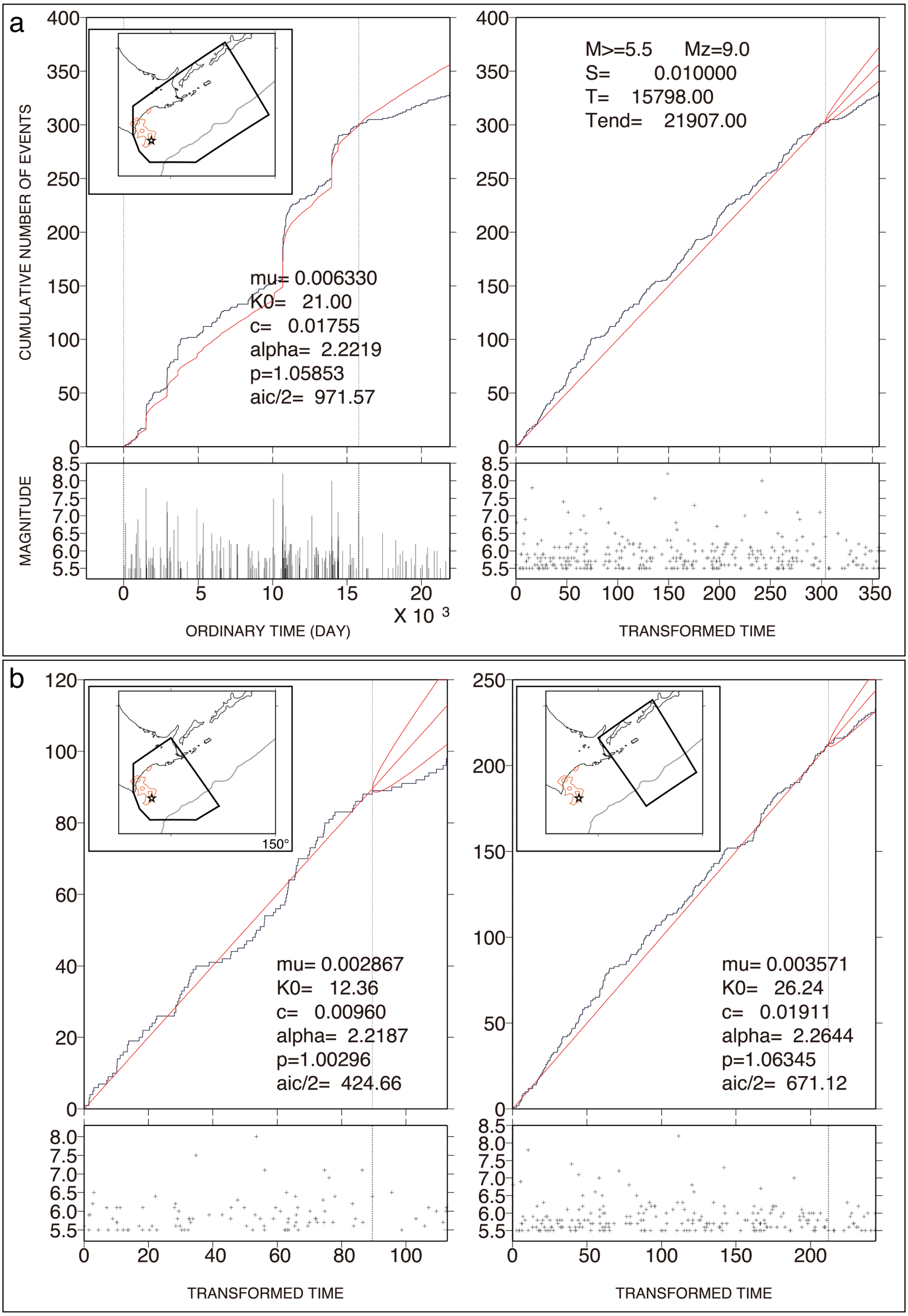

**Supplementary Fig. 7.** Same as Fig. 3 for $M \geq 5.5$.

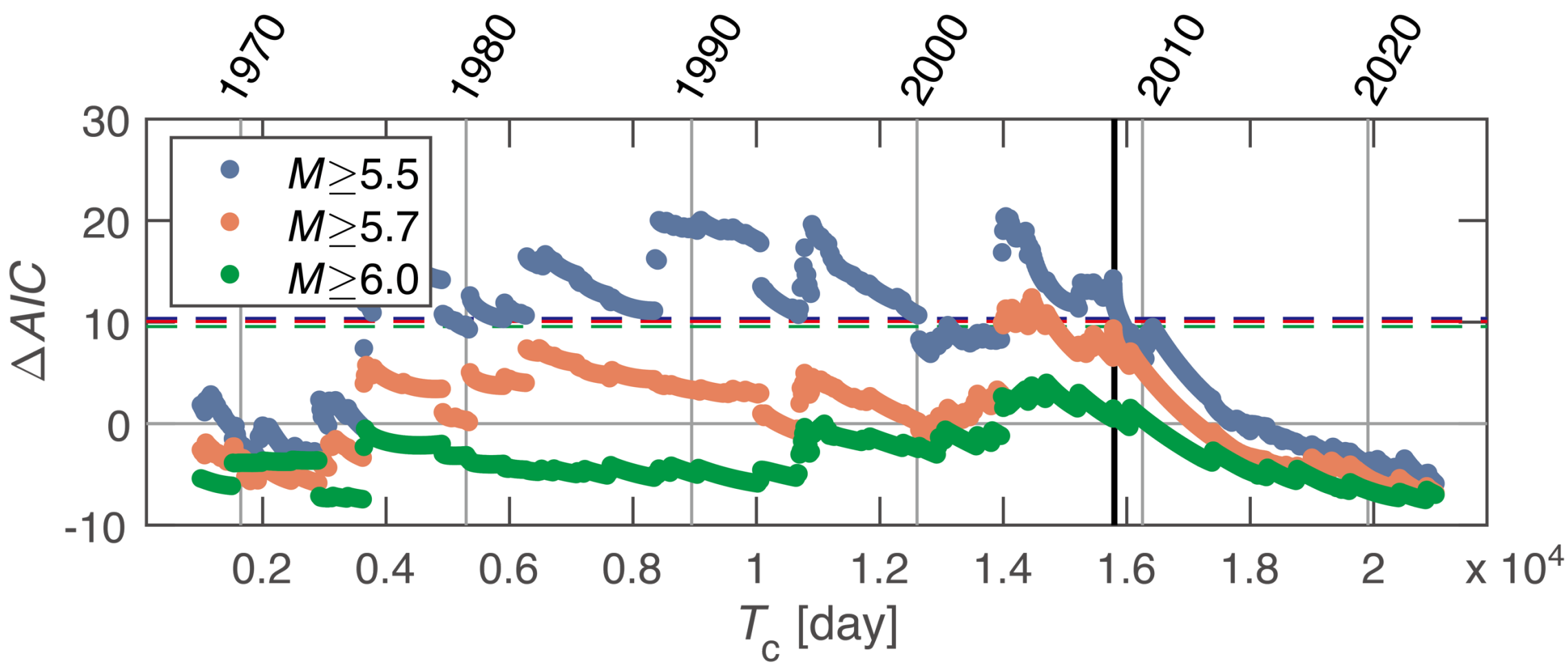

**Supplementary Fig. 8.** $\Delta AIC$ as a function of $T_c$. Blue curve was obtained by using earthquakes with $M \geq 5.5$ ($M_{th}=5.5$) in the region shown in the inset of Fig. 3**a**. Red and green curves are the same as the blue curve for $M_{th}=5.7$ and 6.0, respectively. Horizontal dashed lines representing $2q$ (see Methods for details of searching appropriate $T_c$). As a reference, thin vertical lines indicate Jan. 1 for 1970, 1980, …, 2020. Vertical thick line is Oct. 1, 2008.

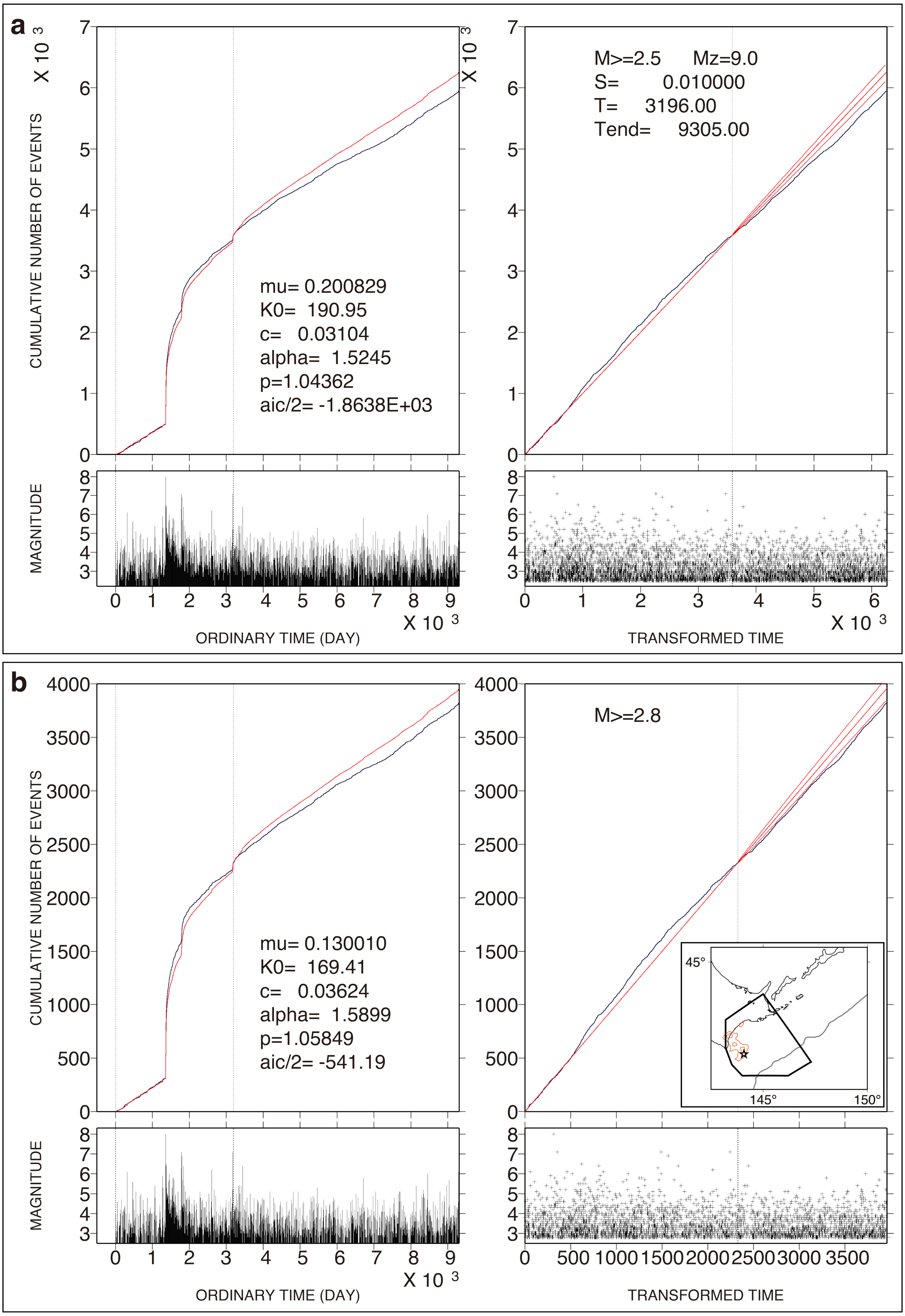

**Supplementary Fig. 9.** Change point analysis for $M_{\mathrm{th}}$=2.5 in **a** and 2.8 in **b** in the region west of the Nemuro peninsula coast (inset of **b**). (**a**) Cumulative function of $M \geq 2.5$

($M_{th}$=2.5) earthquakes is plotted against ordinary time (left panel) and transformed time (right panel), showing the ETAS fitting in the target interval from Jan. 1, 2000 until Oct. 1, 2008 and then extrapolated until Jun. 22, 2025. (**b**) Same as **a** for $M \geq 2.8$ ($M_{th}$=2.8). For details of the figure, see the caption of Fig. 3 and Methods.